%% file: Relax.tex
\documentclass[numberedappendix,iop]{emulateapj}

\usepackage{graphicx}
\usepackage{dcolumn}
\usepackage{bm}
\usepackage{epsf}
\usepackage{verbatim}
\usepackage{amsmath}
\usepackage{amssymb}
\usepackage{appendix}
\usepackage{color}
\usepackage{soul}
\usepackage{verbatim} 
\usepackage{mathtools}
\usepackage{changes}
\usepackage{amsfonts}
\usepackage[percent]{overpic}

\usepackage{xcolor}
\definecolor{darkgreen}{rgb}{0.0,0.5,0.0}
\usepackage[colorlinks,citecolor=darkgreen]{hyperref}

\usepackage{tikz}
\usepackage{mwe}

\usepackage[makeroom]{cancel}

\DeclareMathOperator\arctanh{arctanh}

\input{Definitions.tex}

\def\myfig#1{#1}

\def\specE{s_E}
\def\specP{s_p}
\newcommand{\myD}{\mathcal{D}}

\begin{document}

\title{
Diffusive acceleration in relativistic shocks: particle feedback
}

\author{Yotam Nagar}
\email{nagaryo@post.bgu.ac.il}

\author{Uri Keshet}
\email{ukeshet@bgu.ac.il}

\affil{Physics Department, Ben-Gurion University of the Negev, POB 653, Be'er-Sheva 84105, Israel}

\shorttitle{DSA backreaction}
\shortauthors{Nagar \& Keshet}

\date{\today}

\begin{abstract}
The spectral index $s$ of particles diffusively accelerated in a relativistic shock depends on the unknown angular diffusion function $\mathcal{D}$, which itself depends on the particle distribution function $f$ if acceleration is efficient.
We develop a relaxation code to compute $s$ and $f$ for an arbitrary functional $\mathcal{D}$ that depends on $f$.
A local $\mathcal{D}(f)$ dependence  is motivated and shown, when rising (falling) upstream, to soften (harden) $s$ with respect to the isotropic case, shift the angular distribution towards upstream (downstream) directions, and strengthen (weaken) the particle confinement to the shock; an opposite effect on $s$ is found downstream.
However, variations in $s$ remain modest even when $\mathcal{D}$ is a strong function of $f$, so the standard, isotropic-diffusion results remain approximately applicable unless $\mathcal{D}$ is both highly anisotropic and not a local function of $f$.
A mild, $\sim 0.1$ softening of $s$, in both 2D and 3D, when $\mathcal{D}(f)$ rises sufficiently fast, may be indicated by ab-initio simulations.
\end{abstract}


\maketitle

\section{Introduction}\label{sec:Intro}

Collisionless, non-magnetized, relativistic shocks provide an interesting research target, as they (i) manifest in astronomical systems such as $\gamma$-ray bursts (GRBs);
(ii) may be responsible for ultra-high energy cosmic-rays (CRs);
and (iii) are conceptually simpler than their magnetized or non-relativistic counterparts, being by definition independent of the strength and structure of any preexisting magnetic fields, and independent of the shock Lorentz factor $\gamma_s\gg1$, assuming that the limit $\gamma_s\to \infty$ exists. For recent reviews, see \citet{Sironi2015} and \citet{Pelletier2017}.

Electromagnetic instabilities develop near the shock front, isotropize and thermalize the plasma, and thus induce the shock transition.
The fields generated by the instabilities stochastically scatter a small fraction of highly relativistic particles back and forth across the shock front, thus boosting their energy in a process known as diffusive shock acceleration (DSA).
These non-thermal, so-called CR particles are thought to carry a substantial fraction of the plasma energy, and to modify the electromagnetic fields and the structure of the shock, as seen directly by comparing numerical simulations with and without particle acceleration \citep{Keshet09}.

Resolving the nonlinear interactions between the non-magnetized, relativistic shock, the bulk plasma, the electromagnetic instabilities, and the accelerated CRs, imposes a considerable theoretical challenge.
Studies of the developed shock are based on numerical simulations
\citep[\eg][]{Chang_etal_07, Keshet09,
Nishikawa2009,
Martins2009,
Haugbolle2011,
Sironi2013, Caprioli2014, Caprioli2018, Lemoine2019}
that resolve only the initial stages of the evolution, or on self-similarity assumptions \citep{Katz_etal_07, Medvedev2009} that only gauge the asymptotic scaling laws.

CRs with momenta much higher than any momentum scale in the problem are expected to develop a power-law spectrum.
Working in the limit of small-angle scattering, averaging over planes perpendicular to the shock normal, and assuming that the resulting CR particle distribution function (PDF) $f$ reaches an axisymmetric steady state in the shock frame, $f\simeq q(z,\mu) p^{-\specP}$ can be determined from the similarly-averaged angular diffusion function $\myD\propto D(q;z,\mu)$.
Here, $z$ is the distance from the shock, $\mu\equiv \vect{\hat{z}}\cdot\vect{p} /p$ is the parallel, $z$-component of the CR velocity, $\bm{p}$ is the momentum, and $\specP$ is the momentum spectral index.
However, the functional $D(q;z,\mu)$, and in particular its dependence upon the reduced PDF $q$, are not well understood, and are poorly constrained at the early times probed by ab-initio simulations.

To proceed, one may adopt the test-particle approximation, neglecting the CR backreaction on the electromagnetic field and thus on $\myD$.
If one assumes that $\myD$ is isotropic, an energy spectral index $\specE = \specP-2\simeq 2.22$ can be derived in 3D, numerically \citep{kirk1987, Heavens1988, Kirk_2000, Achterberg2001, Bednarz1998, Ellison2013} and analytically \citep{Keshet_2005}.
Such a spectrum broadly agrees with observations of systems associated with particle acceleration in a relativistic unmagnetized shock, namely GRB afterglows, where $p=2.2\pm0.2$ \citep[][and references therein]{Waxman_06}, and jets in BL-Lac objects, where $p=2.28\pm0.06$ \citep{Hovatta2014}.

For instance, a sample of $\sim300$ GRB afterglows was best-fit by a single value $p\simeq 2.25$, with a broad, $p=2.36\pm0.59$ distribution \citep{Curran2010} probably due to a tail of soft-spectrum systems \citep[\eg][]{Ryan2015}; a similar, $p=2.43^{+0.36}_{-0.28}$ result was found for 38 short GRB afterglows \citep{Fong2015}.
Jets in BL-Lac objects, which show a polarization more consistent with shock acceleration than jets in other active galactic nucleus (AGN) systems, indicate a mean photon spectral index $\alpha=0.64\pm0.03$, corresponding to the aforementioned $p=2.28\pm0.06$, whereas incorporating other types of AGN jets yields a somewhat softer, $\alpha=0.81\pm0.02$ mean spectrum \citep{Hovatta2014}.
The spectrum is not universal among relativistic shocks; an extreme example is the hard spectrum in pulsar wind nebulae \citep[PWNe;][and Arad et al., in prep.]{Fleishman2007}, suggesting a different acceleration mechanism.

While an isotropic $\myD$ yields spectra consistent with the aforementioned observations, $\specE$ was found to be sensitive to the angular dependence of the diffusion function, especially downstream \citep[][and Arad et al., in prep.]{Keshet06}.
Consequently, the CR spectrum is not well-understood in unmagnetized relativistic shocks even in the test-particle approximation, with the peculiar exception of shocks in one dimension \citep[][for a discussion of DSA in an arbitrary dimension, see Lavi et al., in prep., henceforth L20]{Keshet17}. Interestingly, it was recently shown \citep{KeshetEtAl19} that even in non-relativistic shocks, the spectrum is not entirely independent of the diffusion-function anisotropy.

We examine the interrelation between $f$ and $\myD$, which is essential for resolving the DSA problem when particle acceleration is efficient.
We develop a numerical code that solves this problem for an arbitrary ansatz $D(q;z,\mu)$, in both 2D and 3D.
The code is demonstrated for local functionals $D[q(z,\mu)]$, for example a linear relation $D-D_0\propto \pm q(z,\mu)$, where $D_0$ is a constant.
We show that such local relations can arise, for example, if current filaments are created along CR anisotropies, such that their magnetic fields preferentially scatter CRs moving along the filament (plus sign) or temporarily confine them (minus sign).

The paper is organized as follows. In \S\ref{sec:Setup}, we introduce the setup and the transport equations in 2D and in 3D.
Our algorithm and its implementation are outlined in \S\ref{sec:Code}.
In \S\ref{sec:Linear}, we demonstrate the code by investigating the regime in which the functional $D(q)$ is local.
The results are summarized and discussed in \S\ref{sec:Discussion}.
Appendix \S\ref{sec:appendix} presents convergence tests and a discussion of the code parameters.


\section{Setup: DSA in a relativistic shock} \label{sec:Setup}

\begin{figure}[h!]
\hspace*{-0.6cm}
	\centering
\begin{tikzpicture}
    \node[anchor=south west,inner sep=0] at (0,0)
    {\centerline{\epsfxsize=7.5cm \epsfbox{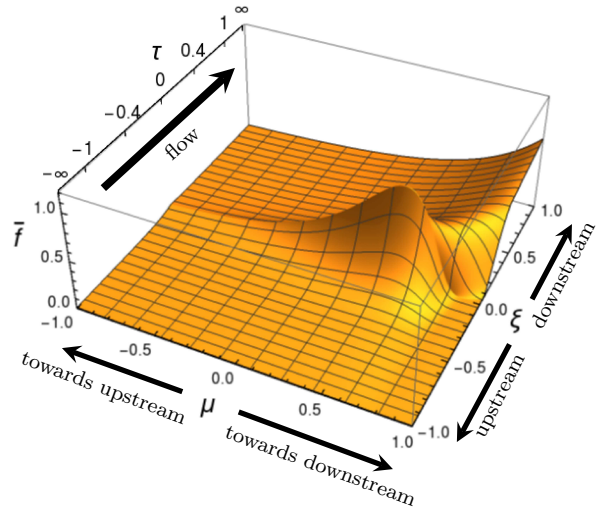}}};
    \draw[line width=2pt,black,-stealth](7.35,1.45)--(6.65,0.15);
    \draw[line width=2pt,black,-stealth](7.65,2.05)--(8.18,3.05);
    \node[label={[label distance=0.5cm,text depth=-1ex,rotate=63]right:upstream}] at (6.5,-0.28) {};
    \node[label={[label distance=0.5cm,text depth=-1ex,rotate=63]right:downstream}] at (7.26,1.17) {};
    \draw[line width=3pt,black,-stealth](1.95,3.528)--(3.7,5.128);
    \node[label={[label distance=0.5cm,text depth=-1ex,rotate=42]right:flow}] at (2.2,3.6) {};
    \draw[line width=2pt,black,-stealth](3,0.82)--(1.4,1.4);
    \node[label={[label distance=0.5cm,text depth=-1ex,rotate=-20.5]right:towards upstream}] at (0.22,1.56) {};
    \draw[line width=2pt,black,-stealth](3.75,0.55)--(5.8,-0.21);
    \node[label={[label distance=0.5cm,text depth=-1ex,rotate=-20]right:towards downstream}] at (2.93,0.57) {};
    \end{tikzpicture}
	\caption{
	Illustration of DSA in 3D by an ultra-relativistic, 
    $\gamma_s\simeq 224$ shock with homogeneous and isotropic angular diffusion, $D=\const$, and (henceforth) the J\"{u}ttner-Synge equation of state.
    The PDF normalized to its maximum, $\bar{f}\equiv f/\fmax$, is plotted in the shock frame, against normalized spatial ($\xi$ or $\tau$) and angular ($\mu$) coordinates. 
	Our relaxation code is used with $N\simeq 1.5\times 10^5$ cells; each rectangle in the plot has $100$ cells in $\xi$ by $5$ cells in $\mu$.
        }
	\label{fig:distribuation_map}
\end{figure}

Consider an infinite, planar shock front at $z=0$, with flow in the positive z direction both upstream $(z<0)$ and downstream $(z>0)$.
Ultra-relativistic particles are assumed to diffuse in momentum angle $\mu$, according to some angular diffusion function $\myD$, leading to a steady-state PDF $f$.
Figure \ref{fig:distribuation_map} demonstrates the shock-frame, normalized PDF for a 3D ultra-relativistic shock with isotropic diffusion.
Here, $z\propto \arctanh(\xi)$ is used to map space onto the compact $-1<\xi<1$ interval.

The figure illustrates the boundary conditions: no particles reach infinitely far upstream, whereas an isotropic (in the fluid frame) PDF $\fiso$ develops far downstream. In the shock frame, $f$ is typically sharply peaked at some forward angle, $\mu>0$, for any $z$. In the upstream (downstream), the decline of $f$ (of $|f-\fiso|$) with increasing distance $|z|$ from the shock is roughly exponential.

The steady-state PDF $f$ of accelerated particles in 3D (see 2D generalization toward the end of this section) satisfies the stationary transport equation \citep{kirk1987}
\begin{equation} c\gamma_i(\beta_i+\tilde{\mu}_i)
\frac{\partial f(z, \tilde{\mu}_i, \tilde{p}_i)}{\partial z} =
\frac{\partial}{\partial \tilde{\mu}_i} \left[
\tilde{\myD}
\frac{\partial f}{\partial \tilde{\mu}_i} \right] \coma
\label{eq:transport1}
\end{equation}
where $\gamma\equiv(1-\beta^2)^{-1/2}$ is the fluid Lorentz factor, $\beta$ is the flow velocity normalized to the speed of light $c$, and $i\in\{u,d\}$ are upstream/downstream indices, written henceforth only when necessary.
A mixed coordinate system is used, with some parameters (designated by a tilde) measured in the fluid frame, and others measured in the shock frame.

The lack of a high energy scale implies a power-law spectrum, $f\simeq \tilde{q}(z, \tilde{\mu})\tilde{p}^{-\specP}$.
The energy spectral index $\specE=\specP-\nu+1$ depends on the momentum spectral index $\specP$ and the number of spatial dimensions $\nu$.
The above boundary conditions now become $\tilde{q}(z\rightarrow -\infty,\tilde{\mu})=0$ upstream, and $\tilde{q}(z\rightarrow +\infty,\tilde{\mu})=\const.$ downstream.
Continuity across the shock front requires $\tilde{q_u}(z=0, \tilde{\mu}_u) \tilde{p}_u^{-\specP}= \tilde{q_d}(z=0, \tilde{\mu}_d)\tilde{p}_d^{-\specP}$, where upstream and downstream quantities are related by a Lorentz boost of velocity $\beta_r=(\beta_u-\beta_d)/(1-\beta_u \beta_d)$.
The velocities $\beta_u$ and $\beta_d$ are related, for a given equation of state, by the Taub adiabat \citep{Taub1948}.
For concreteness, here we adopt the J\"{u}ttner-Synge equation of state
\citep{Synge1957}.

It is customary to assume that the angular dependence of $\myD$ is separable from its energy and spatial behavior.
For our purposes, it suffices to assume that the energy dependence is separable, in the form
$\tilde{\myD}\equiv\left(1-\tilde{\mu}^2\right)\tilde{D}_p\left(\tilde{p}\right)
\tilde{D}\left(z,\tilde{\mu}\right)$.
With the factor $(1-\tilde{\mu}^2)$, $\tilde{D}(z,\tilde{\mu})$ becomes $\tilde{\mu}$-independent for isotropic diffusion.
Rescaling $z$ by defining the optical depth $\tau\equiv z \gamma^3 \tilde{D}_p(\tilde{p})/c$ eliminates $\tilde{D}_p$ from the transport equation.
Boosting all quantities to the shock frame, $p=\gamma \tilde{p}(1+\beta\tilde{\mu})$ and $\mu=(\tilde{\mu}+\beta)/(1+\beta\tilde{\mu})$, one obtains \citep{Keshet06}
\begin{equation}
\mu \frac{\partial q}{\partial \tau} =
\frac{
\partial_{\mu}
  \left\{
  (1-\mu^2)
  D \partial_{\mu}  \left[ \left(1-\beta\mu\right)^{\specP} q \right]\right\}
  }
  {\left(1-\beta\mu\right)^{\specP-3}
  }
\coma
\label{eq:transport2}
\end{equation}
where we defined
\begin{equation}\label{eq:DefQ}
q(\tau,\mu)\equiv\tilde{q}(z,\tilde{\mu})(\tilde{p}/p)^{-\specP}
\end{equation}
and
\begin{equation}\label{eq:DefD}
D(\tau,\mu)\equiv\tilde{D}(z,\tilde{\mu})\fin
\end{equation}

While most DSA studies are in 3D, ab-initio simulations are expensive and are thus often carried out in 2D.
The transport equation (\ref{eq:transport2}) is generalized for the 2D case
in L20.
In analogy with the above, we further modify this equation by separating
$\tilde{\myD}\equiv\tilde{D}\left(\tilde{p}\right)\tilde{D}\left(z,\tilde{\mu}\right)$, and changing  variables to the shock frame.
The transport equation in 2D thus becomes
\begin{equation}
\mu \frac{\partial q}{\partial \tau} =\frac{
\partial_{\varphi}
  \left\{(1-\beta\mu) D
  \partial_{\varphi}  \left[ \left(1-\beta\mu\right)^{\specP} q \right]\right\}
  }
  {
   \left(1-\beta\mu\right)^{\specP-2}
  }
\coma
\label{eq:transport2_2d}
\end{equation}
where $\varphi\equiv\cos^{-1}(\mu)$ is the azimuthal angle, taken here by symmetry in the range $0\leq\varphi\leq\pi$.
The functions $q$ and $D$ are defined in 2D by the same Eqs.~(\ref{eq:DefQ}) and (\ref{eq:DefD}).


\section{
Relaxation code: derive $\!\mbox{\normalsize$\lowercase{q}$}$ given $\!\mbox{\normalsize$D(\lowercase{q;z,\mu})$}$}
\label{sec:Code}

We develop a relaxation code to solve the transport equation (Eq. \ref{eq:transport2} in 3D, or Eq. \ref{eq:transport2_2d} in 2D) for the reduced shock-frame PDF $q(z,\mu)$, given an arbitrary scattering function parametrized as the functional $D(q;z,\mu)$, which we denote $D(q)$ for brevity.
Our nominal algorithm (variants of the method and convergence tests are presented in \S\ref{sec:Discussion} and in Appendix \S\ref{sec:appendix}) may be summarized as follows:
\begin{itemize}
\item Assume some spectral index value $\specP$, map space onto a compact $-1<\xi<1$ interval, and solve the corresponding boundary problem for $q(\xi,\mu)$ using a finite-difference scheme (FDS).
\item Find the value $\specP$ for which the above boundary problem provides the most acceptable physical approximation, by minimizing non-physical negative and oscillatory $q$ behavior.
\item
Start with some $q_0$, repeat the above to construct a series $\{q_j\}_{j=1}^{n}$, where $q_{j}$ corresponds to a diffusion function $D(q_{j-1})$.
Stop when $q_n$ converges, within prescribed accuracy and precision thresholds, on a self-consistent solution $q$ for diffusion $D(q)$.
\item
Repeat the above for incremental grid refinements, and extrapolate the result to an infinite resolution.
\end{itemize}
In the following, we outline each of these steps.

As the spatial variable $\tau$ is unbounded, we change variables to $\xi \equiv \tanh(\tau/\tau_0)$, defined in the finite interval $-1\leq \xi\leq1$.
Increasing the value of the constant $\tau_0$ refines the grid spacing far from the shock (in $z$ or $\tau$ space), and is analogous to strengthening the angular diffusion.
The angular coordinate $-1\leq \mu\leq 1$ is finite, and thus does not require compactification. The transport equation is solved in the Eulerian $\{\xi,\mu\}$ domain.

The transport equation is discretized with a second-order FDS in $\{\xi,\mu\}$ space, using $\{N_\xi,N_\mu\}$ intervals.
The boundary conditions are $q(\xi=-1)=0$ far upstream, and $q(\xi=+1)=(1-\beta_d \mu)^{-\specP}$ far downstream, fixing the overall normalization.
The boundary conditions at $\mu=\pm 1$ are left open (here, the FDS involves a fourth-order, one-sided formula).
This procedure yields $N=N_\xi N_\mu$ linear equations in the $N$ variables $q_{i,j}$, where index $i$ (index $j$) discretizes the coordinate $\xi$ (the angle $\mu$).

The transport equation is second order in $\mu$ and first order in $\xi$, motivating the introduction of an order-unity parameter  $r\equiv N_\xi/N_\mu^2$.
There is considerable freedom in choosing the value of $r$, as well as the value of $\tau_0$, thus modifying the numerical properties of the solution.
For example, for very large $\tau_0$, one cannot resolve the diffusion length near the shock, leading to Gibbs oscillations, whereas a small $\tau_0$ loses resolution far from the shock.
Therefore, for a given shock and diffusion function, we optimize the choice of the parameters $r$ and $\tau_0$ for some small $N$, before proceeding to refine the grid; see discussion in Appendix \S\ref{sec:appendix}.

When discretizing the transport equation (\ref{eq:transport2}) or (\ref{eq:transport2_2d}), the LHS is proportional to $(N_\xi/\tau_0)$, whereas the RHS is proportional to $N_\mu^2$, so it is natural to choose $r\sim \tau_0$; indeed, we typically obtain fast convergence only in this regime.
A notable exception is the case of a non-relativistic shock, $\beta\ll1$. Here, $q$ approaches isotropy, allowing for larger $r$, whereas spatial variations span long, order $\tau\sim \beta^{-1}$ scales, requiring larger $\tau_0$.

A physical solution to the transport equation should be non-negative and non-oscillatory.
The numerical solution for some given, inaccurate, value of $\specP$, typically shows both negative values of $q$ and oscillations along the $\xi$ direction; oscillations in the $\mu$ direction are found to be negligible.
The prevalence of negative $q$ values is quantified using
\begin{equation}
\label{eq:defineQNeg}
\qneg
\equiv
-\frac{\sum_{i,j} \Theta (-q_{i,j}) q_{i,j}}{\epsilon+\sum_{i,j} \Theta (-q_{i,j})} > 0\coma
\end{equation}
where $\Theta$ is the Heaviside step function and $\epsilon>0$ is a small number.
If $q$ were  non-negative everywhere in the grid, then $\qneg$ would vanish, but we have never identified such a behavior.

Oscillations of $q$ in the $\xi$-direction are quantified using
\begin{equation}
\qfluc
\equiv \frac{\sum_{i,j}|q_{i+1,j}-q_{i-1,j}|}{N} > 0\fin
\end{equation}
To estimate the most plausible value $s_0$ of the spectral index $\specP$, we identify $s_0$ as the index minimizing the product of the above two factors, \begin{equation}
    \qeff\equiv \qneg\,\qfluc\fin
\end{equation}

We estimate the uncertainty in the determination of $s_0$ using the variances corresponding to $\qneg$ and $\qfluc$,
\begin{equation}
    \sneg^2 \equiv
    \frac{\sum_{i,j} \Theta (-q_{i,j}) \left(q_{i,j}-\qneg\right)^2}
    {\sum_{i,j} \Theta (-q_{i,j})}
\end{equation}
and
\begin{equation}
    \seff^2\equiv \frac{1}{N}\sum_{i,j}\left(|q_{i,j+1}-q_{i,j-1}|-\qfluc \right)^2
     \coma
\end{equation}
and their weighted mean
\begin{equation}
    \seff=\qeff\sqrt{\left(\frac{\sneg}{\qneg}\right)^2+\left(\frac{\sfluc}{\qfluc}\right)^2}\fin
\end{equation}
The $1\sigma$ confidence interval $\sigma(s_0)$ of $\specP$ is associated with $\specP\sim s_0$ values that satisfy $\qeff(\specP)< \qeff(s_0)+\seff (s_0)$.
The error bars in the following figure correspond to this $1\sigma$ value.

Finally, after repeating the above process for increasingly larger $N$ by refining the grid, we extrapolate $s_0$ and $\sigma(s_0)$ to the physical, $N\to\infty$ limit; see Appendix \ref{sec:appendix} for details.
This scheme provides numerically converged solutions for the spectral index $\specP$ and the PDF $q(z,\mu)$, for any diffusion function $D$. The function $q(\xi,\mu)$ is illustrated in Figure \ref{fig:distribuation_map} for the case of an ultra-relativistic 3D shock with isotropic diffusion, where we obtain $\specP= 4.227\pm 0.001$.

\begin{figure}[!htb]
	\centering
	\myfig{\includegraphics[width=8.5cm]{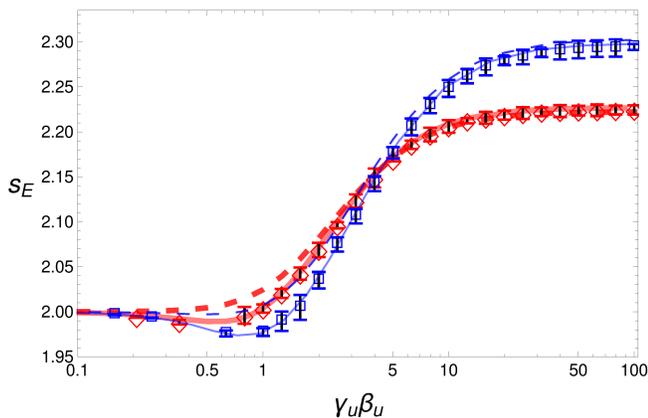}}
	\caption{The energy spectral index $\specE\equiv-d\ln(f)/d\ln(E)$ as a function of the upstream four-velocity $\gamma_u\beta_u$,
	in 3D (red diamonds and thick curves) and in 2D (blue squares and thin curves).
	The results are shown using our relaxation code (symbols, with $2^{16}\leq N\leq 2^{20}$ results extrapolated to $N\to\infty$),  and are compared with semi-analytic methods (solid curves) and with analytic approximations (dashed curves). See text for details.
	}
    \label{fig:iso_velocities_spect}
\end{figure}

Our method reproduces previous results for the behavior of the spectral index, in both 3D and 2D.
This agreement is illustrated in Figure \ref{fig:iso_velocities_spect}, showing $\specE=\specP-2$ in 3D and $\specE=\specP-1$ in 2D, as a function of the shock four-velocity, for a J\"{u}ttner-Synge equation of state and isotropic diffusion.
The results are in excellent agreement with semi-analytic methods: a moment expansion in 3D \citep[][with a Legendre-based expansion of order 7]{Keshet06} and an eigenfunction method in 2D (L20, using an elliptic cosine expansion of order 6).
The results compare favorably with analytic approximations in 3D \citep{Keshet_2005} and in 2D (L20), except in the trans-relativistic regime, where the latter yield slightly softer spectra ($\Delta \specE\lesssim 0.03$) than in our method.

Our code efficiently solves the transport equation for an arbitrarily anisotropic and non-homogeneous $D$.
As long as $D$ is spatially homogeneous on each side of the shock, the problem can also be solved in the moment expansion \citep{Keshet06} method.
We confirm that our code agrees with this method for simple choices of anisotropic but homogeneous choices of $D$.

Importantly, our approach accommodates not only prescribed diffusion functions, but also diffusion functionals $D(q ;\xi,\mu)$ that depend on the PDF $q(\xi,\mu)$ itself.
Such functionals can be used to incorporate the backreaction of the accelerated particles on the scattering electromagnetic modes, as demonstrated in \S\ref{sec:Linear}.
Adding a non-trivial dependence of $D$ upon $q$ substantially complicates the problem, rendering the transport equation non-linear in $q$.
We use an iterative process to solve this non-linear problem.
As the above scheme already computes $q$ for a prescribed $D(\xi,\mu)$, we start with a simple choice of $D$, and then alternate between computing $q(D)$ and computing $D(q)$, iteratively, until convergence is established.

More concretely, we begin with an isotropic and homogeneous diffusion function, $D=D_0\equiv\const.$, henceforth referred to as constant diffusion, on each side of the shock.
We use the above method to find the corresponding PDF, $q_0=q(D_0;\beta_u,\beta_d;N)$, at some resolution $N$. For simplicity of notation, we write this as $q_0=q(D_0)$.
Next, we solve the problem for a modified diffusion function derived from $q_0$, namely $D=D_1\equiv D(q_0)$, and compute the corresponding $q_1=q(D_1)$.
This process is repeated, each step computing $q_k=q[D_k=D(q_{k-1})]$ and minimizing the oscillating and negative $q$ behavior at the same resolution $N$.

The process converges, after some $n$ iterations, onto a well-defined PDF $q_n$ that approximately satisfies $q_n\simeq q[D(q_n)]$, thus providing an approximate solution to the non-linear, numerical transport equation at resolution $N$.
An explicit convergence criterion is defined for terminating the iteration over $k$, namely for deciding when $q_{n}$ is sufficiently close to $q_{n-1}$, for a given resolution $N$:
\begin{equation}
  \frac{1}{N}
      {\sum_{i,j}|q_{n;i,j}-q_{n-1;i,j}|} <
\epsilon \, \mbox{min}
  \left(
    |\bar{q}_n|,|\bar{q}_{n-1}|
  \right)
\coma
\end{equation}
where
\begin{equation}
    \bar{q}\equiv
    \frac{1}{N}\sum_{i,j}q_{i,j}
\end{equation}
is the average value of $q$ over the grid, and $\epsilon$ is a small constant of order $10^{-4}$.

The process is then repeated for increasingly larger $N$, as discussed above, by incrementally refining the grid, until a convergence threshold is met.
The results are then extrapolated to the $N\to\infty$ limit, as discussed in Appendix \ref{sec:appendix}.

\section{Local $D(\lowercase{q})$ dependence} \label{sec:Linear}

Consider an angular diffusion function that is strongly influenced by the accelerated particles.
As a simple demonstration of the code, here we analyze the limit in which feedback by the particles is local in the $\{z,\mu\}$ phase space.
We may further assume, in the limit where particles dominate the plasma evolution, that the fluid frame $\tilde{D}(\tilde{q};z,\mu)\simeq \tilde{D}(\tilde{q}(z,\mu))$ does not depend explicitly on $z$ or on $\mu$.
The physical circumstances under which such simplifications may be justified are discussed in \S\ref{sec:Discussion}.

For simplicity, consider first the case where the functional $\tilde{D}(\tilde{q})$ is linear, such that
\begin{equation}
\label{eq:LinearLocalD}
\tilde{D}(\xi,\mu)
= \tilde{D}(\tilde{q}(\xi,\mu))
=1+ \phi\, \tilde{q}/\qmax
\coma
\end{equation}
where the unit normalization is obtained by rescaling $\tau$.
Here, $-1<\phi<\infty$ is a tunable parameter, which can be either positive (if particles locally enhance the diffusion) or negative (if particles diminish the diffusion), and can be chosen independently upstream and downstream.
The normalization factor $\qmax$, defined as the maximal value of $\tilde{q}$ throughout the respective (upstream or downstream) fluid, is introduced in order to avoid nonphysical, negative values of $\tilde{D}$ for $\phi<0$.
Equation (\ref{eq:LinearLocalD}) is written in the fluid frame,
in order to better describe magnetic structures that move approximately with the fluid; the implied functional form of $D=D(q;\mu)$ in the shock frame then explicitly depends on $\mu$.
Defining Eq.~(\ref{eq:LinearLocalD}) in the shock frame, instead, such that $D=D(q)$ does not explicitly depend on $\mu$, should not qualitatively change our results, as we confirm in several tests.

\begin{figure}[h]
	\centering
	\myfig{\includegraphics[width=8.5cm]{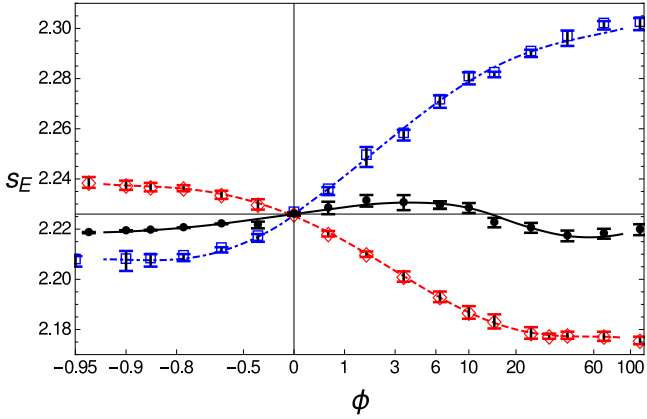}}
	\caption{
	The spectral index $\specE$ in the same  ultra-relativistic shock shown in Figure \ref{fig:distribuation_map}, of $\beta_u = 1-10^{-5}$ ($\gamma_s\simeq 224$) in 3D, with local and linear particle feedback $\tilde{D}(\tilde{q})$ on the angular diffusion function (see Eq.~\ref{eq:LinearLocalD}).
	The horizontal thin line corresponds to a constant (homogeneous and isotropic) diffusion with no feedback.
	Feedback variants demonstrated (symbols, with interpolated curves to guide the eye) include constant diffusion upstream and linear feedback downstream
	($\phi_d=\phi\neq0$; red diamonds),
	the converse --- linear feedback upstream and constant diffusion downstream
	($\phi_u=\phi\neq0$; blue squares), and homogeneous feedback both upstream and downstream ($\phi_u=\phi_d=\phi$; black disks).
	The results are computed using the relaxation code with $N>2^{16}$.
    }
	\label{fig:dasfs}
\end{figure}

The spectrum of particles accelerated by a given shock can be computed as described in \S\ref{sec:Code}, for an arbitrary choice of $\phi_u$ and $\phi_d$, in any dimension.
Figure \ref{fig:dasfs} demonstrates the spectral index $\specE$ obtained in 3D, for an ultra-relativistic shock, with a few simple choices of $\{\phi_u,\phi_d\}$.

\begin{figure*}[!htb]
	\centering
	\myfig{\includegraphics[width=8.5cm]{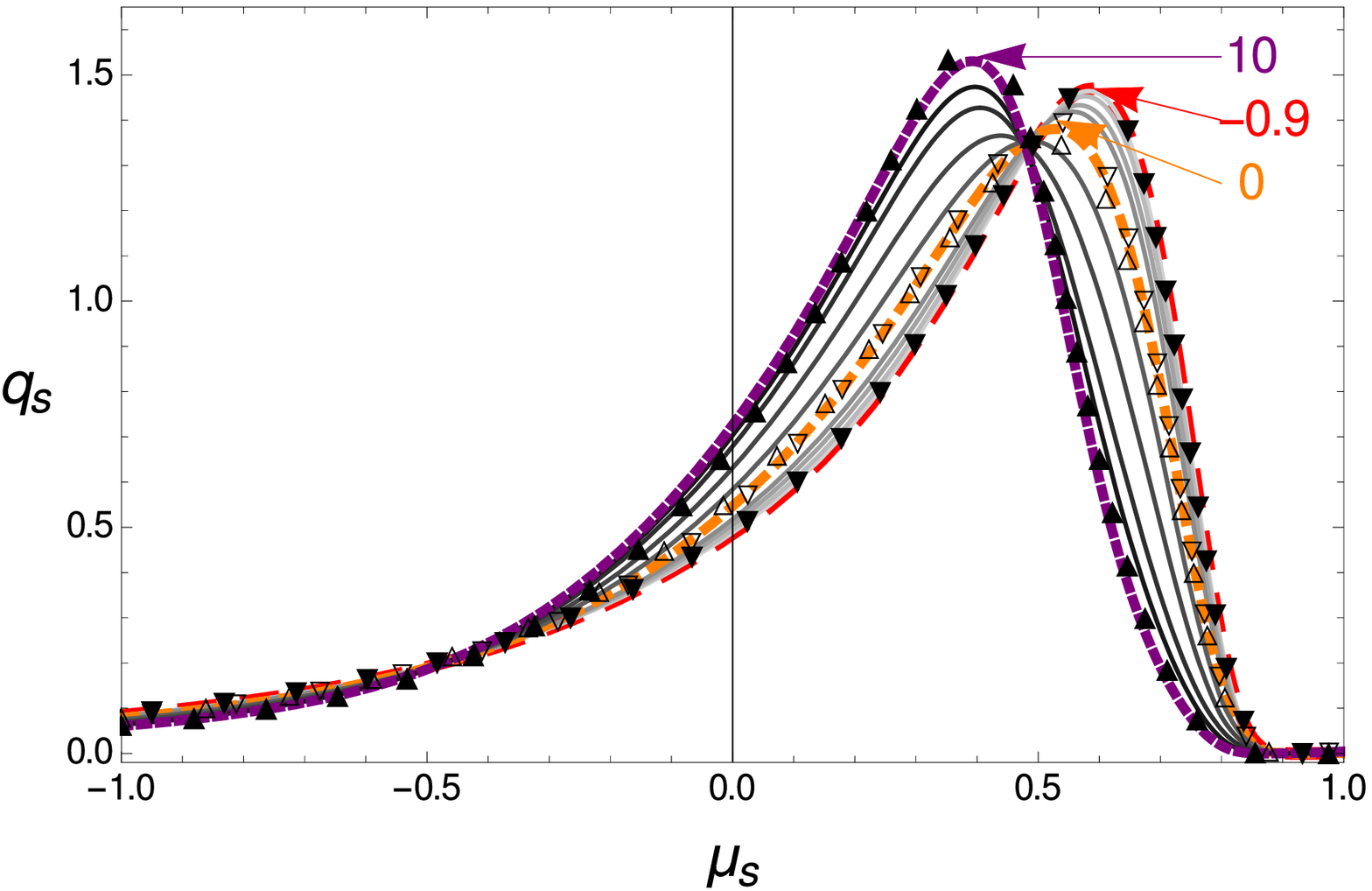}}\hfill
	\myfig{\includegraphics[width=8.5cm]{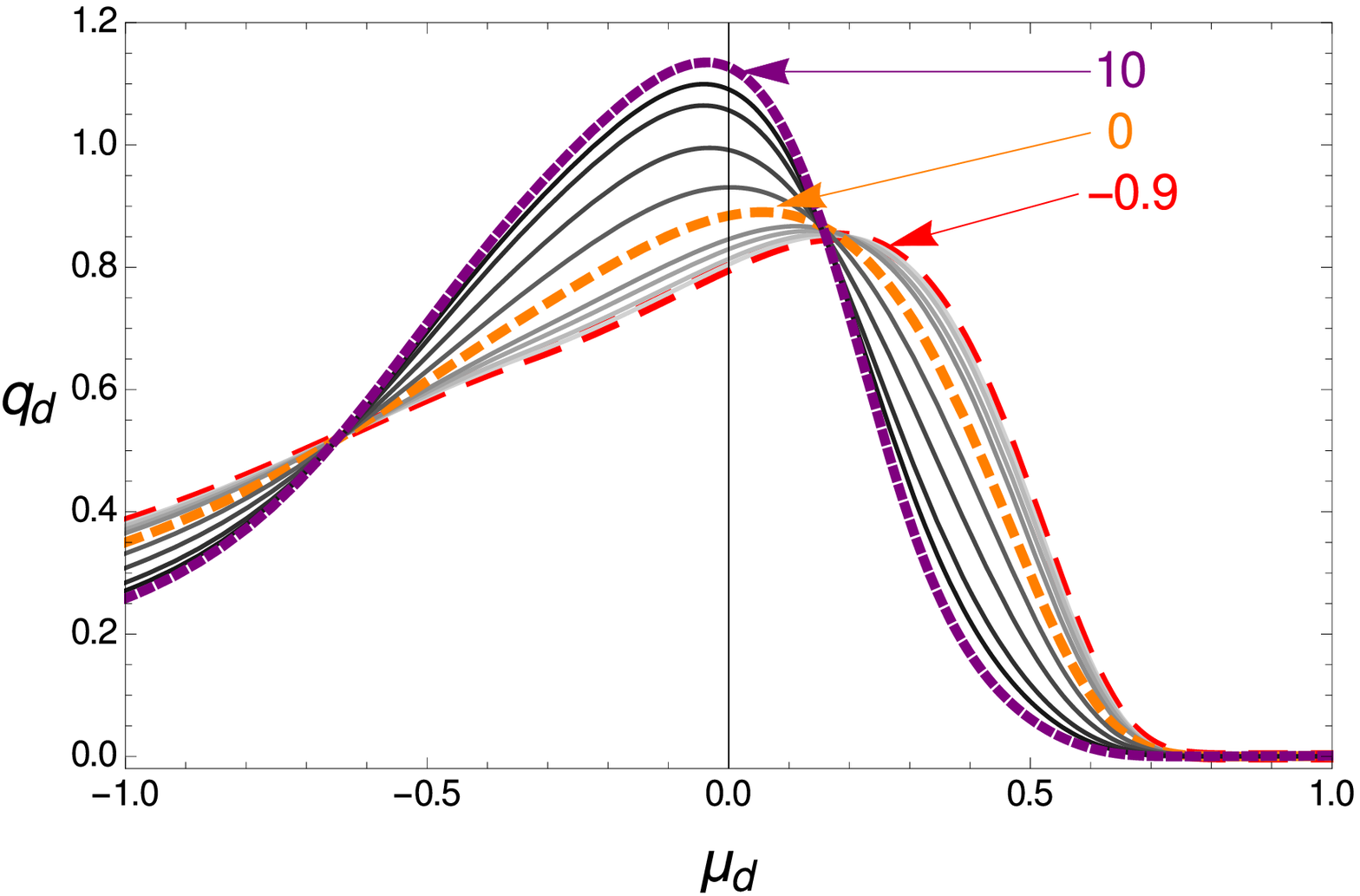}}
	\caption{The angular PDF at the shock front, shown in both shock (left) and downstream (right) frames, computed in the same method and for the same shock as in Figure \ref{fig:dasfs},
	with different choices of linear (see Eq.~\ref{eq:LinearLocalD}) particle feedback.
Several homogeneous choices of $\phi=\phi_u=\phi_d$ are shown (logarithmically spaced between $\phi=-0.86$ and $\phi=6.4$; curves, with a gray scale varying from light to dark as $\phi$ increases).
	Three representative cases, with $\phi\in\{-0.9, 0, 10\}$, are highlighted (annotated dashed curves with increasingly short dashing, colored $\{\mbox{red, orange, purple}\}$).
    Also shown in the shock frame (symbols in the left panel) are the representative cases $\phi=-0.9$ (down-pointing triangles) and $\phi=10$ (up-pointing triangles), applied only to the upstream ($\phi_u=\phi$ and $\phi_d=0$; filled black triangles) or only to the downstream ($\phi_d=\phi$ and $\phi_u=0$; empty triangles).
	}
	\label{fig:distribuation_angular}
\end{figure*}

One test we perform is to vary $\phi_u$ (or $\phi_d$) on one side of the shock, while fixing $\phi_d=0$ (or $\phi_u=0$) for constant diffusion on the other side.
The spectrum is found to become softer (harder) if we choose $\phi>0$ in the upstream (downstream) and $\phi=0$ in the downstream (upstream); the converse is found for $\phi<0$.
For example, we find $\specE \sim 2.3$ ($\specE \sim 2.18$) for $\phi\gg 1$ upstream (downstream), in comparison to $\specE\simeq 2.23$ for $\phi=0$. The upstream effect is slightly stronger than downstream.

Another simple test is to simultaneously vary $\phi_u=\phi_d$, homogeneously, on both sides of the shock.
Here, the opposing effects of particles on each side of the shock roughly cancel each other out, rendering the spectrum approximately unchanged with respect to constant diffusion, within $|\Delta \specE|<0.02$.

These effects are in line with the results obtained in \citet{Keshet06} for local variations in $D(\mu)$.
It was found that enhancing the downstream diffusion at angles $\mu_d\equiv \tilde{\mu}_d\lesssim 0$ ($\mu_d\gtrsim 0$) hardens (softens) the spectrum.
As $\tilde{q}_d$ is largely concentrated for a relativistic shock at $\mu_d\lesssim0$ angles, due to a strong suppression near $\mu_d\simeq 1$ (see Figure \ref{fig:distribuation_angular}), a positive $\phi_d$ (or $\phi_d<0$) yields a harder (or softer) spectrum.
Similarly, the results therein indicate that enhancing the upstream diffusion at angles $\mu_u\equiv\tilde{\mu}_u\lesssim-1+\gamma_u^{-2}$ ($\mu_u\gtrsim-1+\gamma_u^{-2}$) softens (hardens) the spectrum. As $\tilde{q}_u$ is strongly concentrated for a relativistic shock at $\mu_u<-\beta_u$, a positive $\phi_u$ (or $\phi_u<0$) yields a softer (or harder) spectrum.
Local changes in $D$ have a somewhat stronger effect in the downstream \citep{Keshet06}, but as $\tilde{q}_u$ is far less isotropic than $\tilde{q}_d$, the weighted effect of changing $\phi$ is slightly stronger upstream.

\begin{figure}[!htb]
	\centering
	\myfig{\includegraphics[width=8.5cm]{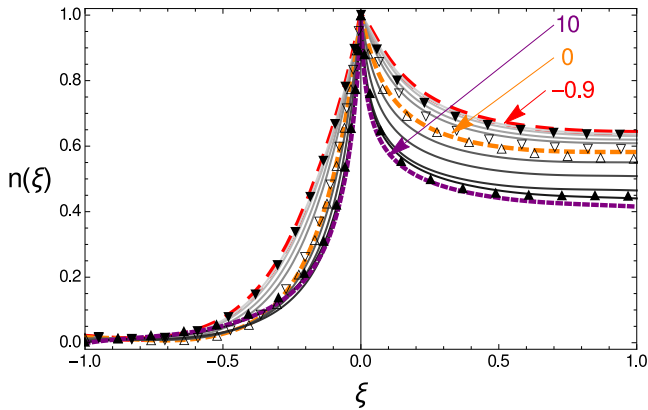}}
	\caption{
	Spatial distribution of shock-frame particle density, $n(\xi)$, normalized to its value at the shock front, computed in the same method and for the same shock as in Figures \ref{fig:dasfs} and \ref{fig:distribuation_angular}, for different choices of linear (see Eq. \ref{eq:LinearLocalD}) feedback, with the same notations as in Figure \ref{fig:distribuation_angular}.
	}
	\label{fig:distribuation_spatial}
\end{figure}

The dependence of the PDF $q(\xi,\mu)$ upon the choice of the feedback parameters $\phi_u$ and $\phi_d$ is illustrated in Figures \ref{fig:distribuation_angular} and \ref{fig:distribuation_spatial}.
Figure \ref{fig:distribuation_angular} shows the angular distribution at the shock front, measured both in the shock frame and in the downstream frame.
Figure \ref{fig:distribuation_spatial} presents the spatial evolution of the particles, parametrized in terms of the surface density measured in the shock frame,
\begin{equation}
n(\xi)\equiv \int_{-1}^{1} q(\xi,\mu) \,d\mu \fin
\end{equation}
These figures show the PDF obtained both for a uniform, $\phi_u=\phi_d=\phi$ feedback (curves), and for feedback on only one side of the shock (symbols).

Figure \ref{fig:distribuation_angular} shows that an increasingly positive (negative), uniform $\tilde{D}(\tilde{q})$ feedback, \ie $\phi=\phi_u=\phi_d>0$ ($\phi<0$), gradually shifts the angular distribution towards the upstream (downstream) direction.
Figure \ref{fig:distribuation_spatial} shows that a positive uniform feedback, $\phi>0$, confines the particles closer to the shock, whereas $\phi<0$ leads to a more spatially extended distribution.

While the reduced PDF $q(\xi,\mu)$ depends on the $\tilde{D}(\tilde{q})$ feedback upstream, it is approximately independent of feedback downstream.
The empty symbols in Figures \ref{fig:distribuation_angular} and \ref{fig:distribuation_spatial} demonstrate that for $\phi_u=0$, similar distributions are obtained for any choice of $\phi_d$.
In contrast, the filled symbols indicate that the choice of $\phi_u$ does modify $q$, and furthermore, a modified $\phi_u\neq 0$ distribution remains insensitive to $\phi_d$.
We verify that the PDF variations due to changes in $\phi_d$, while modest, suffice to self-consistently explain the corresponding small changes in $\specE$, by confirming the same spectrum directly from the energy gain and escape probability inferred from $q(\xi=0,\mu)$.

These results, their higher sensitivity to $\phi_u$ with respect to $\phi_d$, and the stronger effect of large positive $\phi$ values with respect to negative $\phi$, can be qualitatively understood by inspecting the transport equation (\ref{eq:transport1}).
For this purpose, one may use, for example, the first upstream eigenfunction, which well-approximates the PDF for isotropic scattering \citep{Kirk_2000}.
Evaluating the RHS of Eq.~(\ref{eq:transport1}) at the shock front for each term in the diffusion function (\ref{eq:LinearLocalD}), $\tilde{D}=1$ and $\tilde{D}\propto \tilde{q}$, separately, indicates that the corresponding contributions to $\pr_\tau \tilde{q}(\tilde{\mu})$ are quite similar to each other downstream, up to a normalization, but are different upstream.
Similarly, $\tilde{D}=1+\phi \,\tilde{q}/\tilde{q}_{max}$ yields $\pr_\tau \tilde{q}(\tilde{\mu})$ that is approximately constant, up to a normalization, for $\phi\to-1$, but changes as $\phi>0$ increases.
Consequently, modifying $\phi_d$ or $\phi<0$ only weakly affects the solution.

Overall, as illustrated by Figure \ref{fig:dasfs}, we generally find that a local feedback by the accelerated particles on the diffusion function has a rather modest effect on the spectrum.
This behavior is not limited to the linear form of the $\tilde{D}(\tilde{q})$ functional adopted in Eq.~(\ref{eq:LinearLocalD}).
The deviation of the spectrum from the constant-diffusion case is found to remain modest, even if one invokes stronger local dependencies, including quadratic ($\tilde{D}-1\propto \tilde{q}^2$) and cubic ($\tilde{D}-1\propto \tilde{q}^3$) functionals.

\section{Summary and discussion}
\label{sec:Discussion}

We have developed a relaxation code for studying particle acceleration in a general planar, arbitrarily relativistic shock, in both 2D and 3D, in the small-angle scattering limit.
Unlike previous studies, we solve the problem for an arbitrary angular diffusion function $D$ that depends not only on the particle angle $\mu$ and location $z$, but is also an arbitrary functional of the PDF $f$.
The code is based on a finite difference scheme, iteratively relaxed to self-consistently solve for the spectral index $\specE$ and the reduced  PDF $q(\xi,\mu)$, where $\xi=\tanh(\tau/\tau_0)$ is the rescaled optical depth.
The code can be easily generalized for multiple dimensions and large-angle scattering.

The known solutions for the simple case of isotropic diffusion are reproduced; Figures \ref{fig:distribuation_map} and \ref{fig:iso_velocities_spect} demonstrate the resulting $q(\xi,\mu)$ and $\specE$, respectively, for an ultra-relativistic shock in 3D.
To illustrate the more general scenario, in which $D$ depends explicitly on $f$, we study the simple case where this dependence is local in $\xi$ and $\mu$, as exemplified by Eq.~(\ref{eq:LinearLocalD}) for a linear $\tilde{D}(\tilde{q})$ dependence.
As shown in Figures \ref{fig:dasfs}--\ref{fig:distribuation_spatial}, a positive (negative) proportionality coefficient $\phi_u$ in the upstream softens (hardens) the spectrum, shifts $f$ toward upstream-directed, more negative $\mu$ angles (toward downstream-directed, large positive $\mu$), and leads to stronger (weaker) confinement of particles to the shock front.
In the downstream, $f$ is more isotropic; although $\phi_d>0$ ($\phi_d<0$) leads to a slightly harder (softer) spectrum, it does not have a significant impact on the PDF.

We vary the analysis by considering different local, fluid frame prescriptions for $\tilde{D}(\tau,\tilde{\mu})=\tilde{D}(\tilde{q}(\tau,\tilde{\mu}))$, such as quadratic or cubic $\tilde{D}(\tilde{q})$ functionals.
In all cases, we obtain roughly similar results, both for the spectral index $\specE$, and for its qualitative dependence upon $\phi\propto d\tilde{D}/d\tilde{q}$.

Overall, a local particle feedback on the diffusion function does not substantially alter the spectral index, which remains within $|\Delta \specE|\lesssim 0.1$ from its value for isotropic diffusion, as demonstrated in Figure \ref{fig:dasfs}.
For example, the most extreme change we find in $\specE$ for diffusion functions of the form (\ref{eq:LinearLocalD}), is obtained when $\phi_u\to\infty$.
For an ultra-relativistic shock in 3D, the spectral index then converges on $\specE \simeq 2.3$, in comparison to $\specE\simeq 2.2$ for isotropic diffusion.
The spectrum for this $\tilde{D}\propto \tilde{q}$ behavior is softer than found for quadratic or cubic $\tilde{D}(\tilde{q})$ scaling.

We deduce that the spectral index is reasonably well-approximated by its isotropic-diffusion value, unless $\tilde{D}$ is both highly anisotropic and not a local function of $\tilde{q}$.
Our conclusions pertain to both 3D and 2D.
Simple analytic approximations for the spectral index in 3D \citep{Keshet_2005} and 2D (L20) are thus applicable, within $|\Delta \specE|\lesssim 0.1$, for such local particle feedback.

The phase space of functionals $\myD(f(\bm{r},\bm{p},t);\bm{r},\bm{p},t)$, in general non-local, which one should consider around a given shock, is vast. Our code addresses functionals of lower dimension, $D(q;\tau,\mu)$, involved in the effective $1+1$ dimensional formulation of the steady-state problem when a power-law spectrum is applicable and the energy dependence of $\myD$ is separable. Moreover, the local functionals $\tilde{D}(\tilde{q})$ analyzed above, in the limit where particle feedback is strong, depend on $\tau$ and $\mu$ only through $\tilde{q}(\tau,\mu)$, and thus form a small subset of plausible diffusion functionals.
This example, presented as a proof of concept for our code, can be easily generalized for more realistic diffusion functionals.
While the practical relevance of the present example to astronomical shocks is unclear, there are several reasons why such simplified local functional may be useful as a toy model.

First, as particle acceleration is thought to be efficient, CRs should carry a non-negligible fraction of the energy, and so should affect $\myD$.
It is natural to expect long wavelength electromagnetic modes, capable of efficiently scattering high energy CRs, to be driven by similarly energetic CRs \citep[\eg][]{Katz_etal_07}.
Such a behavior is confirmed by particle in cell (PIC) simulations, showing that particle acceleration is essential for the generation of long wavelength modes \citep{Keshet09}.

Assuming that CRs thus influence the scattering modes responsible for their own angular diffusion, and given the short range of the interaction with these modes, electromagnetic mode generation by CRs would contribute to a spatially-local dependence of $\tilde{D}$ upon $\tilde{q}$. Averaging out spatial dimensions perpendicular to the flow, the remaining dependence of $\tilde{D}$ upon $\tilde{q}$ may be approximated as local in $z$.
Such locality can, however, be in part smeared by the advection of modes with the flow.

It is more difficult to see to what extent $\tilde{D}(\tilde{q})$ should be local in $\tilde{\mu}$.
In general, a highly anisotropic PDF will generate instabilities, that tap on the free energy of the distribution and isotropize it.
If the deviation from anisotropy is localized around a given direction, some instabilities preferentially scatter the particles moving in nearby directions.
Under such circumstances, and inasmuch as the small-angle scattering is valid, the influence of CRs on $\tilde{D}$ in a given direction will be correlated with $\tilde{q}$ in the same direction.
In the strong correlation limit, and after averaging out the azimuthal direction, the contribution of $\tilde{q}$ to $\tilde{D}$ becomes approximately local in $\tilde{\mu}$.

Consider for example the Weibel-like electromagnetic instabilities \citep{Weibel1959, 1959Fried}, generated due to the streaming of relativistic particles in some $\pm \unit{v}$ direction.
Such instabilities lead to current filamentation, that in the linear stage generates magnetic fields with a long (short) coherence length parallel (perpendicular) to $\unit{v}$.
Due to their large Larmor radii, the relativistic particles are affected mainly by the long coherence-length modes.
Consequently, such an instability induces stronger diffusion parallel to the same direction $\unit{v}$ in which the PDF was enhances, resulting in a positive correlation between the diffusion function and the PDF.
A similarly local, but negative, correlation between $\tilde{D}$ and $\tilde{f}$ can also develop, for example if the long wavelength modes parallel to $\unit{v}$ become sufficiently strong and extended to confine the relativistic particles.

Both behaviors can manifest upstream of a weakly magnetized shock, due to Weibel-like instabilities \citep{Gruzinov_Waxman_99, Achterberg2007}, as well as oblique two-stream instabilities \citep{Bret2008, Bret2009, Nakar2011},
which show in the wave frame a similarly longer coherence length parallel to the flow \citep{Plotnikov2013}.
Simulations of ultra-relativistic, weakly magnetized shocks show strong filamentation in the shock precursor, where the energetic particles are beamed into a narrow, $1+\tilde{\mu}\lesssim\gamma_u^{-2}$ cone opposite to the flow direction.
The resulting magnetic filaments are elongated along the flow, and are seen to gradually grow in scale \citep{Keshet09, Sironi2013, Lemoine2019}. The resulting electromagnetic structures enhance the diffusion of highly relativistic particles parallel to the flow, and increasingly trap intermediate-energy particles.

We conclude that while a local $\tilde{D}(\tilde{q})$ particle feedback is an over-simplification, it may be relevant to streaming shock-accelerated particles.
Interestingly, the spectral index inferred from ab-initio simulations is typically somewhat softer than anticipated theoretically for isotropic diffusion.
For example, 2D PIC simulations find spectral indices in the range $\specE \simeq 2.3\mbox{--} 2.5$ \citep{Spitkovsky2008, Sironi2013}, whereas the spectrum anticipated in an ultra-relativistic shock in 2D is $\specE\simeq(1+\sqrt{13}/2)\simeq 2.30$ for isotropic diffusion (L20).
We find that sufficiently strong $\tilde{D}_u(\tilde{q}_u)$ correlations soften the 2D spectrum to $\specE\simeq 2.4$, consistent with the spectrum in ab-initio simulations, suggesting a possible feedback effect in the latter.

\acknowledgments

We thank A. Spitkovsky and I. Gurwich for helpful discussions.
This research has received funding from the GIF (Grant No. I-1362-303.7 / 2016), and was supported by the IAEC-UPBC joint research foundation (Grants No. 257/14 and 300/18), and by the Israel Science Foundation (Grant No. 1769/15).

\appendix
\section{Convergence and sensitivity tests} \label{sec:appendix}

\begin{figure}[!htb]
\myfig{\includegraphics[width=9.5cm]{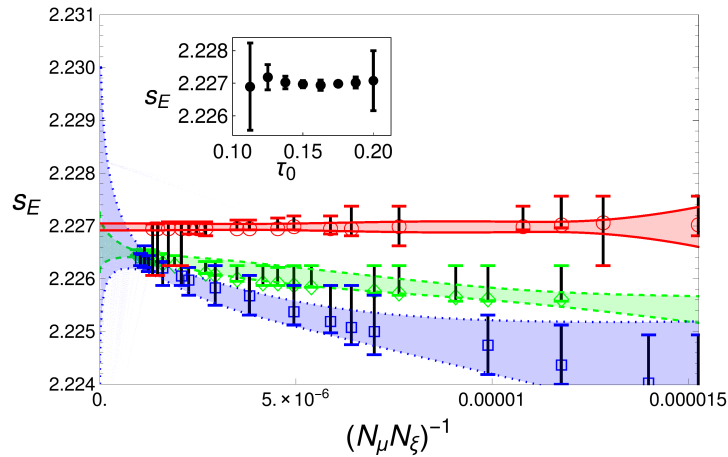}}
\centering
	\caption{
    Convergence plot of $\specE$ for the ultra-relativistic shock, with constant diffusion, shown in Figure \ref{fig:distribuation_map}.
 The spectral index (symbols with $1\sigma$ error bars) is shown as a function of the number of cells, along with a power-law fit ($1\sigma$ shaded region between curves), for three different choices of the resolution ratio, $r\equiv N_\xi/N_\mu^2=0.5$ (blue, squares, dotted curves), $1$ (red, circles, solid), and $4$ (greed, diamonds, dashed), showing best convergence for $r\simeq 1$.
 Inset: extrapolated $\specE$ in the $N\rightarrow\infty$ limit for $r=1$, with different choices of $\tau_0$, showing best convergence for $\tau\simeq 0.15$.
 }
 	\label{fig:68bands}
\end{figure}

Our code uses $N= N_\xi N_\mu$ grid cells, where $N_\xi$ is chosen even to avoid cells split in half by the shock.
The results are extrapolated to the physical, $N\rightarrow \infty$ limit using a power-law fit of the form $\specE=s_0+c N^\alpha$, where $s_0$, $c$, and $\alpha\leq 0$ are fit parameters.
Figure \ref{fig:68bands} demonstrates the convergence for $2^{16}\leq N\leq 2^{20}$, along with the $1\sigma$ confidence interval of the fit, for an ultra-relativistic shock in 3D with constant diffusion.
We use the Pearson's chi-square test to quantify the goodness of fit, and demand a p-value smaller than $0.01$.

The figure shows that while the results are converged, the rate of convergence and the uncertainty in the extrapolation depend somewhat on the free parameters of the code.
In particular, the convergence rate depends on the resolution ratio $r\equiv N_\xi/N_\mu^2$, and on the spatial rescaling parameter $\tau_0$, defined through $\xi \equiv \tanh(\tau/\tau_0)$.
The figure shows results for three different choices of $r$, and the inset demonstrates different choice of $\tau_0$ for $r=1$.

For a given shock and diffusion function, our code estimates the optimal values of $r$ and $\tau_0$ by sampling a few small values of $N$, and minimizing the error in $\specE$ extrapolated to $N\to\infty$.
Typically, for a 3D relativistic shock as shown in Figure \ref{fig:68bands}, the optimal free parameters are $r\simeq 1$ and $\tau_0\simeq 0.15$.
Namely, such an ultra-relativistic shock benefits from comparable resolutions ($r\simeq 1$) per differential order in the transport equation, and an enhanced resolution near the shock by a factor $\tau_0^{-1}\simeq 7$.

For non-relativistic shocks, larger values of $\tau_0$ are needed to better resolve the long, $\tau\sim \beta^{-1}$ spatial diffusion scale, and larger values of $r$ are needed to enhance the spatial resolution with respect to the angular resolution, because the latter becomes redundant as the anisotropy decreases.
Variations in the shock parameters are somewhat sensitive to $\phi$, and become more substantial in 2D shocks.

Convergence is found to be robust, as verified by testing different code variations.
In particular, different discretization schemes were tested, and found to converge on the same results; a second order FDS in $\{\tau,\mu\}$ space was found to be optimal.
Different starting points $D_0$ for the relaxation process were tested, and found to converge on the same results.
No bifurcation as a function of $r$ and $\tau_0$ was identified.

\vspace{0.5cm}
\bibliography{DSA}
\bibliographystyle{apj}

\end{document}

%% file: Definitions.tex
\newcommand{\ie}{\emph{i.e.} }
\newcommand{\eg}{\emph{e.g.,} }

\newcommand{\be}{\begin{equation}}
\newcommand{\ee}{\end{equation}}
\newcommand{\bea}{\begin{equation*}}
\newcommand{\eea}{\end{equation*}}
\newcommand{\beqr}{\begin{eqnarray} \nonumber}
\newcommand{\eeqr}{\end{eqnarray}}
\newcommand{\beqrb}{\begin{eqnarray}}
\newcommand{\eeqrb}{\nonumber \end{eqnarray}}
\newcommand{\fin}{\mbox{ .}}
\newcommand{\coma}{\mbox{ ,}}

\newcommand{\const}{\mbox{const}}

\newcommand{\vect}[1]{\mathbf{#1}}
\newcommand{\unit}[1]{\hat{\mathbf{#1}}}
\newcommand{\pr}{\partial}

\def\seff{\sigma_{\mbox{\scriptsize eff}}}
\def\sfluc{\sigma_{\mbox{\scriptsize fluc}}}
\def\sneg{\sigma_{\mbox{\scriptsize neg}}}
\def\qmax{\tilde{q}_{\mbox{\scriptsize max}}}
\def\fmax{f_{\mbox{\scriptsize max}}}
\def\fiso{f_{\mbox{\scriptsize iso}}}

\def\qneg{\bar{q}_{\mbox{\scriptsize neg}}}

\def\qfluc{\bar{q}_{\mbox{\scriptsize fluc}}}
\def\qeff{\bar{q}_{\mbox{\scriptsize eff}}}

